# Controlled Transformation of Electrical, Magnetic and Optical Material Properties by Ion Beams

**B.A. Gurovich , D.I. Dolgy, E.A. Kuleshova, E.P. Velikhov, E.D. Ol'shansky, A.G. Domantovsky, B.A. Aronzon, E.Z. Meilikhov**

**Russian Research Center "Kurchatov Institute",123182 Moscow, Russia**

## Summary.

Key circumstance of radical progress for technology of XXI century is the development of a technique which provides controllable producing three-dimensional patterns incorporating regions of nanometer sizes and required physical and chemical properties. Our paper for the first time proposes the method of purposeful direct transformation of the most important substance physical properties, such as electrical, magnetic, optical and others by controllable modification of solid state atomic constitution.

The basis of the new technology is discovered by us effect of selective atom removing out of thin di- and polyatomic films by beams of accelerated particles. Potentials of that technique have been investigated and confirmed by our numerous experiments. It has been shown, particularly, that selective atom removing allows to transform in a controllable way insulators into metals, non-magnetics into magnetics, to change radically optical features and some other properties of materials.

The opportunity to remove selectively atoms of a certain sort out of solid state compounds is, as such, of great interest in creating technology associated primarily with needs of nanoelectronics as well as many other "nano-problems" of XXI century.

## Introduction.

Traditional methods of efficient modification of substance physical properties are based on chemical processes resulting in principal changing combination of physical properties of final products as compared with those of interacting initial reagents. Analogous transformation of properties is possible under chemical decomposition of di- or polyatomic compounds. Similar processes can be accompanied by the efficient modification of the electrical conductivity, magnetic or optical properties and other physical and chemical qualities of a material, and have long been used in practice. In addition to chemical reactions leading to efficient changing atomic composition of substances, there is another method associated with nuclear reactions of different kinds. The first of these methods (chemical one) has so wide field of applications that those could not even be enumerated. As for the second one, the field of its uses is quite narrow and the latter one is of exotic character often.

It is essential that no one of those methods can provide *controlled spatial modulation* of the transformation process of atomic composition in the body volume (first of all, in solids). In other words, those methods do not allow to produce a *given bulk "portrait"* of local atomic composition variations in *controllable* way.

Developing such a method could result in producing materials with controlled spatial modulation of needed physical and chemical properties. Something like this is realised now in producing metal layers in microprocessors. However, in that case the result is reached not by means of changing atomic composition but with the help of complex multiple-stage physicochemical (and even mechanical) treatment of different materials consecutively used.

Our work demonstrates the possibility of the purposeful efficient changing atomic composition of solids by the action of beams consisting of accelerated particles of certain energies. That transformation of atomic composition is not associated with some chemical or nuclear reactions but is connected with the *selective removing atoms of a given sort* out of di- or polyatomic compounds as a result of atomic displacements by the beam of accelerated particles. Similar variations of atomic composition can lead to efficient changing physical properties of substances and, in particular, to the transformation of insulators in metals and non-magnetic materials in magnetic ones, to modification of optical properties and so on.

## 1. Physical basis of selective removing atoms of a given sort out of di- or polyatomic compounds with the help of atomic displacements

Let us consider the situation which arises under the interaction of monoenergetic beam of non-relativistic particles of energy $E$ and mass $m$ with diatomic crystals consisting of atoms with masses $M_1$ and $M_2$. Maximum energy transferred to the atoms of the solid is equal [1]

$$E_{\max}^{(1,2)} = \frac{4mM_{1,2}}{(M_{1,2}+m)^2}E, \qquad (1)$$

where $E_{\max}^{(1)}$ and $E_{\max}^{(2)}$ are maximum energies which accelerated particles could transfer to atoms with masses $M_1$ or $M_2$. The displacement of atoms out of crystal lattice points with generating a stable Frenkel pair is a threshold effect and occurs in the case when the energy transferred to atoms is more than the threshold displacement energy $E_d$[#]. [2]. If the crystal consists of atoms of the only sort, $E_d$ has the only value for each crystallographic direction. As a rule, $E_d$ - value is of a few tens of eV that is an order of magnitude higher than the sublimation energy. Such a high value could be explained as follows. To generate stable Frenkel pair, extracted interstitial atom has to be removed from the vacancy over a distance ~ 5 lattice periods. During this transfer, the interstitial atom interacts with crystal atoms positioned close to its trajectory. That needs a certain energy input depending on the number of atoms which interact with the interstitial one in the course of its transfer from the own vacancy to the nearest stable position and on the coordination of those atoms which is defined by the trajectory of the interstitial atom relative to some lattice crystallographic directions.

From the preceding it is clear that threshold energies for atoms of different sorts in di- or polyatomic crystals are, in general, diverse for each crystallographic direction. In addition, it follows from the relationship (1) that under the irradiation of di- or polyatomic crystals, maximum energies transferred to atoms of different kinds by accelerated particles are diverse. The stronger is the difference of atomic masses, the higher is that distinction. If $m<M_1<M_2$ then $E_{\max1} > E_{\max2}$, and if $M_1<M_2<m$ then $E_{\max1} < E_{\max2}$.

[#] In some cases, the generation of Frenkel pairs (radiation defects) under irradiation could be associated not only with threshold effects which are specific for atom displacement out of lattice points due to *elastic* (quasi-elastic) scattering of incident particles. In insulators, Frenkel pairs could arise through exciting electron subsystem of a crystal due to *non-elastic* interaction of incident particles with solid state atoms. That phenomenon is well known for wide-gap ion insulators of which typical examples are alkali-halide crystals [3]. In the present work that effect is not considered as: 1) in every cases, classic threshold generation of radiation defects have been observed with all studied compounds; 2) it is *not possible* to produce *directed displacement* of interstitial atoms by means of Frenkel pair generation due to excitation of electron subsystem in solids.

Therefore, varying the mass and the energy of beam particles one could attain conditions when the higher energy would be transferred to lighter (if $m<M_1<M_2$) or heavier (if $M_1<M_2<m$) atoms in di- or poly-atomic crystal. That opens up new possibility of selective removing only light (or only heavy) atoms out of di- or polyatomic crystal. Such selective removing atoms of a single sort under the beam action is possible in the case when the maximum energy transferred to atoms exceeds the threshold energy $E_d$ for atoms of that sort only.

Given above considerations concerning displacements of different atoms in *crystals* are applicable, in great part, to the same compounds in *amorphous state*. For the problem considered, the most significant distinction of amorphous materials from crystals is that there is no spatial anisotropy of displacement threshold energy in amorphous substances. However, that distinction is not an obstacle for selective removing atoms, if above-mentioned conditions are fulfilled.

Thus, it is clear that for the beam incident normally to the crystal surface conditions could be built up when removing atoms of a single sort in the direction of the incident beam occurs at the crystal thickness comparable with the projective range of beam particles in di- or polyatomic crystal. In doing so, other atoms would not be subjected to any directed transfer. It allows to diminish significantly the concentration of selected atoms in the corresponding crystal layer or remove them completely by varying the fluence ($\Phi_t$) of incident particles. As a result, one could cause efficient modification of physical and chemical properties in relevant crystal layers or thin films. Such a transformation can occur in layers which thickness is comparable with the projective range of beam particles in the irradiated material.

Some obvious features of the considered mechanism of selective removing atoms may be formulated as follows:

1. The rate of selective atoms removing is proportional to the flux density of incident beam particles.
2. The process of selective atoms removing under the action of the incident beam is, in nature, non-thermal one in a wide range of irradiation temperatures that distinguishes this process principally from chemical reactions and processes of substance transformations according to phase diagrams.
3. The process of selective atoms removing out of crystals could be realized through the upper (relative to the beam) additional layer of different material if its thickness is lower than the projective range of beam particles in this layer. If, in addition, the threshold energy of atom displacement in the additional layer is higher than maximum energy transferred to them from beam particles, then the directed displacement of atoms in that material does not occur. In the opposite case, penetrating atoms of the material in the underlying layer and their transfer in the beam direction could occur over the distance comparable with the projective range of beam particles in the sandwich considered.

Above-mentioned features of the proposed method of action on thin films or layers indicate its perspective for the efficient, purposeful and spatially-modulated modification of composition, structure, physical and chemical properties of materials.

## 2. Modification of structure, electrical, magnetic and optical properties by selective atoms removing.

In deciding on a particular material for selective atoms removing, metal compounds that in the initial state are insulators hold the greatest practical interest. Among the diatomic compounds, those are, for example, many metal oxides, as well as some hydrides and nitrides of metals. Though in the course of this work experiments were performed with compounds of all above-mentioned types, metal oxides were investigated most thoroughly. Qualitative features of effects accompanying selective atoms removing are identical in compounds of all above-mentioned types. The aim of experiments was to remove selectively oxygen (nitrogen or hydrogen) atoms by irradiation of the original insulator and to obtain finally a metal. Experiments have been performed with thin films of different thickness, which have been produced by reactive sputtering of metals in the atmosphere of relevant gases (oxygen, nitrogen or hydrogen). [4]

In most cases, films have been irradiated by protons with energies from ~150 up to ~1200 eV. In addition, in some experiments films have been irradiated in electron microscope column by electrons with energy of 100 – 200 keV.

In the work, complex investigations of initial and irradiated films were performed which included: measurement of electrical resistance within the temperature range 4.2-300 K, magnetic and optical properties[4], structural measurements by means of transmission electron microscopy and electron diffraction analysis[5], tunnel microscopy technique[6] as well as by methods of X-ray photoelectron spectroscopy[7].

To measure the electrical resistance of films in the course of the proton irradiation, special through electrical contacts were built in the film substrates. One of the contact ends was polished abreast with the surface of the substrate on which studied film was deposited later.

Films were deposited on substrates of different insulators with high specific resistance (higher than $10^9$ Ohm·cm), for instance, on diamond-like coat or glass. Reference experiments with irradiating substrates themselves show that radiation-induced effects of changing their resistance are not significant as compared with observed effects of changing properties of studied films.

Film investigations of different insulating di- and polyatomic materials show that their behaviour under irradiation depends strongly on the energy of beam protons, all other factors being the same. Observed radiation-induced modification of thin film properties was of clearly defined threshold character[##]. No modifications of structure, composition, electrical and magnetic properties of thin films were detected until the proton energy reached certain minimum value (which is individual for every of studied material). According to the preceding, it means that as long as the energy $E_{\max}$ transferred to material atoms by protons is low ($E_{\max} < E_{d1}$, $E_{d1}$ where $E_{d1}$, $E_{d1}$ are displacement threshold energies for atoms of the first and the second types, respectively), atom displacements out of lattice points do not take place. With increasing proton energy, conditions appear when $E_{d1} \leq E_{\max} < E_{d2}$. Then principal variations of structure, composition, electrical, magnetic and optical material properties begin to be observed. As shown below, all those variations are connected with selective removing atoms of oxygen, nitrogen or hydrogen out of studied compounds (oxides, nitrides or hydrides, respectively). With further increasing proton energy, the condition $E_{d1} \leq E_{d2} < E_{\max}$ is reached and atoms of both types in diatomic compounds begin to displace out of lattice points. Within relevant energy diapason, selectivity of atom displacements diminishes with increasing $E_{\max}$. Such a regime is almost not considered below as being less interesting.

Now we consider in detail experimental results on selective atoms removing for di-atomic compounds in the regime when the condition $E_{d1} \leq E_{\max} < E_{d2}$ is fulfilled. Investigations show that initial insulating films of different materials are in one of three structural states:

1. Polycrystal state with grains of 10-100 nm;

[##] In what follows, basic effects are predominately demonstrated for diatomic oxides (because of their essential qualitative similarity for all compounds).

2. Amorphous state;
3. Combined state with variously oriented grains distributed in amorphous matrix or separated by thin amorphous layers.

Electron diffraction patterns of all above-mentioned types of films correspond to the annular diffraction specific for polycrystals or look like a diffuse halo or, at last, appear to be a composition of both those types of diffraction.

It should be also noticed that the most of studied films have a crystal structures which do not correspond to the handbook data for the same stable-state compounds. It is known that similar situation is characteristic for thin films and associated with their intrinsic inclination for polymorphism and nonequilibrium phases formation [8]; the same could lead to different anomalies in thin film properties.

Experiments show that, independently of the structure state of original insulators, they behave equally in the course of selective atoms removing. As a rule at the first stage of irradiating insulators (of polycrystalic or combined structure) amorphization of crystallites takes place. That effect is detected clearly by disappearing crystal contrast in the dark-field electron-microscopic images (see Fig. 1a). With this, significant reducing of the intensity and broadening of point reflexes (lines) in annular electron diffraction patterns (up to their complete transformation in diffuse halo characteristic to amorphous materials) are observed (see Fig. 1 b,c).

Further irradiation is accompanied by yet another phase transformation corresponding to the transition from amorphous state to crystal one. This is evident from appearing crystal contrast in the dark-field electron-microscopic images associated with newly generated grains (see Fig. 1a). Simultaneously, absolute or partial (depending on original material) disappearance of diffuse halo, increasing intensity and appearing new system of diffraction rings in electron diffraction patterns occur, that testify appearing a new (metallic) phase (see Fig. 1c,d). The formation of a metallic phase after irradiation with a high fluence is demonstrated below when presenting investigation results.

At the most cases, diffraction in crystallites of newly generated metallic phase does not correspond to those types of crystal lattices which are cited in handbooks for relevant pure bulk metals. At the same time, diffraction in crystallites of new metallic phase (produced from insulators due to selective atoms removing) in some instances corresponds to diffraction data which have been obtained for thin films of relevant pure metals. For example, with selective removing of oxygen atoms out of oxide $WO_3$, metallic tungsten has been generated in FCC- lattice of period $a$=4.19±0.02 Å instead of BCC lattice typical for bulk tungsten. As mentioned above, that could be associated with polymorphism characteristic for thin films. However, in some works (see, for instance[9]) formation of thin W-films with FCC lattice of period $a$=4.15 Å have been observed as a result of pure tungsten sputtering. Taking into account that those authors[9] have estimated the accuracy of their electron diffraction measurements within ~ 2%, one should consider the coincidence of lattice parameters as very good. Nevertheless, in some cases metal films been produced by selective atoms removing have got crystal lattice identical to that of the same bulk metal. For example, copper produced out of oxide CuO demonstrates the diffraction characteristic of FCC-lattice with parameter $a$=3.60±0.02 Å while the handbook value is $a$=3.615 Å.

Experiments show that variation of irradiation temperature (from –198 up to 300 $^{o}$C) does not influence the rate of phase transformations (amorphization and subsequent crystallisation).

In spite of some distinctions, dependencies of electrical resistance of different films on proton fluence have common features. As a rule, at the initial stage of selective removing atoms of the certain kind (for instance, oxygen atoms) sharp reducing (up to ~$10^{10}$ times) of the film resistance is observed (see Fig. 2).

However, for compounds with low starter specific resistivity, temporary increasing of specific resistivity is at first observed during the initial stage of irradiation with subsequent sharp reducing of it (cf. Fig. 3). Further, with increasing irradiation fluence from ~10 up to ~ 400 dpa###, the specific resistivity value continues to slowly reduce asymptotically approaching a limit value. The latter is defined by properties of the metal which oxide is subjected to irradiation. Eventually, irradiation reduces the specific resistivity up to a factor of $10^{12}$ at room temperature.

In the proton energy range $E_{d1} \leq E_{\max} < E_{d2}$, outlooks of dose specific resistivity dependences are quite similar qualitatively and quantitatively for equivalent irradiation times.

At higher proton energies, when $E_{\max}>E_{d2}$, the character of dose resistance dependencies is changed due to displacements of metal atoms (cf. Fig. 2, curve 2,3). In the case when $(E_{\max}-E_{d2})\leq$3-5 eV, only weak increasing resistance of film material is observed at high proton fluences (cf. Fig. 2, curve 2). But if transferred and threshold energies differ significantly enough ($E_{\max}-E_{d2}\geq$50-80 eV), the character of dose resistance dependence is changed sharply (cf. Fig. 2, curve 3).

X-ray photoelectron spectroscopy of the film material irradiated by proton beams of different fluences shows that increasing fluence of protons which energy corresponds to the condition $E_{d1} \leq E_{\max} < E_{d2}$ is accompanied by monotonous reducing oxygen in film material (see Fig. 4). After proton irradiation with high fluence (~100-400 dpa for oxygen atoms), there are only traces of oxygen in those films. With this, specific resistivity of the films takes values typical for metal films (~$10^{-3}$-$10^{-5}$ Ohm·cm) but distinctly exceeding handbook values for bulk materials. Thus, the energy $E_{d1}$ corresponds to the threshold displacement energy of oxygen atoms in studied di-atomic compounds. It should be noticed that for investigated oxides this energy varies over wide limits: from 35 eV up to ~100 eV (in some cases this value is more than 100 eV).

Dose dependencies represented in Fig. 2 demonstrate that the energy $E_{d2}$ corresponds to the threshold displacement energy of metal atoms. Increasing electrical resistance value at high proton fluences when $E_{\max}>E_{d2}$ is conditioned by reducing film thickness due to atom displacement which could be readily detected experimentally. Threshold displacement energy of metal atoms for oxides investigated amounts from ~20 eV up to ~70 eV.

To ensure that the conductivity of materials obtained under irradiation is of the metal type, we have studied resistance temperature dependencies of different materials. In Fig. 5 shown are temperature dependencies of the conductivity for: 1) original CoO-oxide film, 2)the same film irradiated with two different proton fluences and also 3)the film of the same thickness obtained by ion sputtering of pure metal. (For the sake of clearness, dependencies of the reduced resistance *R(T)/R*(300 K) are represented in the figure).

It is seen that with enlarging dose the temperature resistivity coefficient *dR/dT* increases and even changes its sign from the negative one, typical for insulators (curves 1, 2 in Fig. 5), to positive one (curves 3, 4 in Fig. 5) corresponding to the metal conductivity. For metal films obtained by selective atoms removing, extrapolation of temperature resistance dependencies to $T$=0 results in clearly nonzero residual conductivity that unambiguously indicates the metal type of their conductivity. More detailed analysis shows that resistance of original insulator and that of the material obtained at low irradiation fluences are well described by the relationship $\rho \propto \exp[(T_0/T)^{1/2}]$ known for hopping conductivity where parameter $T_0$ diminishes with increasing irradiation dose. That dependence is well known and describes the conductivity of systems where charge transfer occurs via electron tunneling. With further increasing irradiation dose, in

### Abbreviation "dpa" means Displacement Per Atom.

most cases the conductivity temperature dependence of studied materials develops an appearance typical for metal glasses or amorphous and quasi-amorphous metals and looks like a similar dependence for a deposited film of pure metal (cf. Fig. 6 and the insert in it). Indeed, at relatively high temperatures the temperature resistivity coefficient is positive and dependence $R(T)$ is approximately linear, at lower temperatures there is a characteristic minimum on the $R(T)$-curve (cf. Fig. 6), and at even lower temperatures $R \propto \ln T$ that is in a good agreement with known results for metallic glasses[10].

In some cases (for example, copper produced out of oxide CuO) this dependence is practically monotonous and looks like that for metal film produced by ion sputtering of pure copper (Fig.6, curves 2,4).

Apart from the above-cited features, some characteristic peculiarities would be noted which confirm the resemblance of obtained materials to quasi-amorphous (ultradisperse) metals. Firstly, characteristic values of the residual resistivity for obtained materials and quasi-amorphous metals are similar[10] and much higher than those of crystal analogs. Secondly, in both cases temperature resistivity coefficient values are on order of several percent per Kelvin[10]. At last, there is a weak dependence of the conductivity on the film thickness associated with the significant role of electron scattering on intergrain boundaries, within crystal grains and in amorphous phase. (Thickness dependence of film conductivity is discussed below in more details.) Estimates of electron free path $l$ in the material give a value of ~ $(0.3–10)\cdot 10^{-7}$ cm, so in the case considered $l$~10 Å <<$d$ ~ 100 Å that excludes manifestation of the classical size effects[11]. Irradiating the mostly refractory metal - tungsten, we did not succeed in obtaining positive temperature resistivity coefficient, but even in that case there was a nonzero residual conductivity (specific resistivity is ~$8\cdot 10^{-4}$ Ohm·cm) and, hence, a metallic type of conductivity of produced material are of no doubt. At helium temperatures the sign of temperature resistivity coefficient becomes positive and experimental results could be described by the relationship $\sigma=1/\rho \propto \ln T$ characteristic for quasi-two-dimensional disordered metals[12].

*Table 1.* **Specific resistivity of metal films (μOhm·cm) produced by various methods as a function of their thickness ($T_{measure}$=20°C).**

| Material | Film thikness, Å | | | | |
|---|---|---|---|---|---|
| | *50* | *100* | *200* | *500* | *1000* |
| Cu* | 44 | 14 | 9.3 | 7.4 | 6.2 |
| Cu** | -- | -- | 19 | -- | -- |
| Cu*** | 1.68 | | | | |
| W* | 163 | 133 | -- | 105 | 65 |
| W** | -- | 800 | -- | 960 | 700 |
| W*** | 5.39 | | | | |
| Co** | 163 | 161 | 159 | -- | -- |
| Co*** | 6.24 | | | | |
| Fe* | -- | 35 | 26 | -- | |
| Fe** | -- | 360 | -- | -- | -- |
| Fe*** | 9.72 | | | | |
| Al* | 134 | 16.9 | -- | 11.1 | 10.1 |
| Al*** | 2.73 | | | | |

* - metal obtained by reactive ion sputtering of pure metal
** - metal produced by selective removing of oxygen atoms out of oxide
***- standard value for bulk metal

Experiments demonstrate that with varying thickness $d$ of irradiated films over a wide range (but, naturally, with $d$ not exceeding the projective proton range in the relevant material), the resistance $R$ of metal films (obtained under identical irradiation conditions) changes accordingly to the relationship $R \propto \rho/d$ where ρ is the specific resistivity of the film material.

In Table 1 are shown results of specific resistivity measurements for films of metals produced by ion sputtering of pure metal targets and the same metals obtained by selective removing of oxygen atoms out of oxide with proton irradiation. For comparison, standard specific resistivity values for bulk metals are presented.

Results presented in Table 1 show that specific resistivity of thin metal films are strongly dependent of their thickness. As this takes place, the smaller film thickness, the greater distinctions from standard values for bulk samples. For films of 10 nm-thickness, those distinctions are on order of value, and for films of 5 nm-thickness they are even more. As has been mentioned above, it is, probably, associated with the classic size effect [11]. There is no escape from the notice that in metal films generated by selective oxygen atoms removed out of oxides (in the same thickness range 10-100 nm) specific resistivity practically does not dependent on the thickness (see Table 1). As has been noticed, it is associated with a small value of electron free path as compared with the film thickness. Electron-microscopic investigations demonstrate that significant distinctions of specific resistivity for films of minimum thickness generated out of pure metals are most likely connected with the amorphous component in film structure. In such films (especially, in the thinnest of them) volumes of amorphous and crystalline phases are comparable that is clear from corresponding electron diffraction patterns. With increasing thickness of pure metal films, the volume part of amorphous phase diminishes sharply and for the thickness of 50-100 nm it disappears completely. Simultaneously with lowering amorphous phase, enlarging of the mean crystallite size in pure metal films is observed. Effects of thickness influence on the film pure metals structure are most likely conditioned by difficulty of grains' growth in the course of deposition due to surface proximity that promotes conservation of amorphous phase in films. Besides, it is known that crystallite growth in thin films depends strongly on conditions of their condensation (firstly, on the temperature and the rate of their deposition) [4,8]. There is a different situation if one obtains metal films by selective atoms removing out of insulators - grains' growth is conditioned by crystallite generation from of amorphous state during phase transition. Our experiments show that the mean crystallite size and the existence (or absence) of amorphous component do not depend on the thickness of irradiated insulator films. There is a clear tendency: the lower is melting temperature of the metal produced by film of insulator irradiation, the larger is the crystalline size (and, respectively, the smaller is the part of amorphous phase) and the lower is the distinction between measured specific resistivity and standard one (see table 1). The same reason could probably explain distinctions between temperature dependencies of conductivity for films produced on the base of Cu, Co and W. The effect of melting temperature on specific resistivity and its temperature dependence, crystallite size and disappearance of amorphous phase is mostly evident for copper produced by selective removing oxygen atoms out of oxide. Copper has the lowest melting temperature among all studied metals. That is, likely, the reason for the similarity of all its properties to that of copper films produced by ion sputtering of pure metal.

Experimental results concerning film thickness variation (at a given proton energy) evidence that the projective range $l_p$ of protons with the energy of ~ 1 keV equals ~100 nm for $WO_3$-film. When the thickness of films irradiated by protons of mentioned energy becomes more than $l_p$, measured resistance turns to be higher than that calculated with the relationship $R \propto 1/d$ assuming that transformation of insulator to metal occurs over the whole film thickness.

It appears to be of interest to investigate the influence of proton flux on the rate of resistance variation during selective atoms removing out of oxides and their transformation in metal. Experiments demonstrate that the rate of insulator transformation in metal increases proportionally to the proton flux. That effect is readily seen under analysis of dose dependencies of resistance obtained with different proton fluxes for the same materials.

Special-purpose experiments evidence possibility of selective atoms removing out of insulator (with transforming them in metal) also in the case when its surface presented to the beam of accelerated particles is covered by an additional thin layer of different material. Those experiments have been performed in two, principally distinct, variants. In the first one, the insulator has been covered by additional thin film of a substance which atoms have been displaced in the course of irradiation just as atoms selectively removing out of insulator. As it is evident from dose specific resistivity dependencies of Fig. 7, in that case insulator transforms eventually in metal in the same way as without of additional layer between the beam and insulator (compare Fig. 7 and Fig. 2). However, there are some peculiarities at initial stages of dose conductivity dependencies connected with transferring atoms of additional layer through the insulator layer under the action of proton beam (see Fig. 7). The existence of atoms of additional layer in transformed insulator (carbon atoms, in the case considered) leads to certain increase of its resistivity and so promotes lowering the rate of insulator transformation into metal. Increasing the thickness of additional layer prolongs the transformation time (compare Fig. 7, curves 1, 2). However, additional layer practically does not influence the final resistivity of the metal film produced out of insulator. It occurs because under high proton fluence additional layer disappears completely - in part, due to sputtering but, essentially, due to migration of layer atoms through the insulator/metal film. Eventually, additional layer atoms are removed out of insulator/metal to the substrate. In order that process takes place, several conditions have to be met: proton fluence has to be high enough and, in addition, the projective range of beam ions has to be comparable with the total thickness of irradiated sandwich.

In the second variant of the experiment, additional rhenium layer has been used which atoms have not been removed by beam protons of a given energy (due to large rhenium mass). That has been proved in special preliminary experiments in which rhenium films have been irradiated by protons under the same conditions and rhenium resistance appears to be unchanged. The thickness of additional rhenium layer has been selected so that its resistance $R$ being 9 times higher than that of Co-film ($r_\infty$) produced by selective removing oxygen atoms out of cobalt oxide under high irradiation dose (without additional rhenium layer) at the same geometry, i.e. $R=9r_\infty$.

Let $r(t)$ is a dose dependence of oxide film resistance without rhenium film in the course of proton irradiation under conditions considered. In a test experiment, the third of oxide film surface (between contacts to measure resistance) has been covered by additional rhenium layer. At the early irradiation stage, when $r(t)>>R$, resistance $R_s$ of the sandwich with rhenium layer equals $R_s=(1/3)R+(2/3)r(t)\approx(2/3)r(t)$. If the additional rhenium layer prevents selective atoms removing out of oxide, then under high dose $R_s$ could be represented as the resistance of the film part not covered by rhenium $(2/3)r_\infty$ connected consecutively in series with the resistance of rhenium part. In the case, as mentioned above, the latter equals $3r_\infty$ and is well below the resistance of oxide film lying beneath the rhenium layer. Thus, $R_s=(2/3)r_\infty+3r_\infty=(11/3)r_\infty$. If the additional rhenium layer does not effect selective removing atoms out of the insulator, then under high irradiation dose $R_s=(2/3)r_\infty+(1/3)r_\infty\cdot 3r_\infty/[(1/3)r_\infty+3r_\infty]=(29/30)r_\infty$ . Therefore, under high irradiation dose, sandwich resistances have to differ by a factor of about 3.8 for two considered cases.

Experiments shows that under irradiation of such a sandwich up to high dose, its measured resistance agrees with the latter relationship with an accuracy of ~15-20%. Hence, the conclusion could be made that the additional rhenium layer does not prevent oxide transformation into metal by means of selective atoms removing and does not effect the resistance value of the metal film reached after irradiation with high proton fluences.

One could also expect that with selective removing oxygen atoms out of initially nonmagnetic (or weak magnetic) oxides of ferromagnetic metals, films could transform into magnetic state. It is clear that in relevant experiments other di- or polyatomic compounds of ferromagnetic metals could be employed along with oxides.

Below represented are results of our experiments performed with certain di- and polyatomic systems.

Magnetisation of films deposited on nonmagnetic substrates has been measured by a special magnetometer designed for thin film measurements. Selective atoms' removing has been induced by proton irradiation of different energies. In addition to dose resistance dependencies (cited above), in the course of those experiments threshold displacement energies $E_{d1}$, $E_{d2}$ and projective ranges for protons of selected energies have been also defined.

Investigations of insulator films CoO, $Fe_2O_3$ and Fe-Co-V-O show that original states of two first of them are nonmagnetic (see, for example. Fig. 8, curve 1) and the third film has a weak magnetisation (Fig. 8, curve 3). In bulk, those oxides are weak ferromagnetics. However, as has been mentioned above, the structure of thin films produced by reactive sputtering is distinctive of bulk sample structure. Probably, the same reason leads to the difference between magnetic features of films and bulk materials.

After irradiation by protons with energies meeting requirements $E_{d1} \leq E_{max} < E_{d2}$ (for each of oxides above-mentioned), transformations of all those insulators into respective metals have resulted (Fig. 9). Subsequent measurements have shown that all metals generated as a result of selective removing oxygen atoms out of oxides are ferromagnetics (Fig. 8, curve 2,4). Thus, in the case of CoO и $Fe_2O_3$ oxides selective atoms removing induces not only the transformation of the insulator to metal but the transition of non-magnetic material into ferromagnetic one, as well. In the case of Fe-Co-V-O oxide, irradiation results in increasing of saturation magnetisation by an order of value. It should be noticed that the specific magnetisation of metal films produced in such a way is several times (1.5-3) lower than standard values for the bulk materials or thin films obtained by ion sputtering of respective metals. Reasons for that are likely the same as those which lead to the difference between specific resistivity of bulk materials and that of the same metals produced by selective atoms removing out of insulators.

During the process of selective atoms removing out of various compounds, the density of a material can change. It is clear that such a process has to result in increasing the material density (as compared with original one) due to bulk relaxation of metal atoms into voids appearing after removing of oxygen atoms out of oxides (or some different atoms out of compounds of other types).

That assumption has been verified by special experiments which have been performed in a following way. The surface of insulator film has been covered by a mask with regular transparent holes and then the film has been irradiated by protons with energies providing selective atoms removing out of the insulator. Hence, open sections of the film have been irradiated only. Proton fluence is high enough and provides producing transformed metal films with the resistance close to minimum one. After irradiation, the relief generated at the film surface has been studied with the help of tunnel microscope. Measurements show that thickness diminishing of irradiated section is of 20-50 % (depending on the chemical composition of insulators). It is essential that in spite of such a significant film thickness diminishing, there are no changes of linear film dimensions in the film plane. That conclusion is confirmed by the following fact: in our numerous experiments, we never observed neither film exfoliation nor violation of its continuity. This fact is especially wonderful as the transformation into metal by selective atoms removing passes amorphization stage. For this reason, variations of linear dimensions leading to changing of film volume would be isotropic. Observed strong anisotropy of linear dimensions' variations (with diminishing film volume after irradiation) could be explained as follows. With diminishing film volume, mechanical stresses arise on the film-substrate inter-phase boundary. Impossibility of film deformations in the boundary plane results in tensile stresses which act on the film and could diminish film linear dimensions (that is to deform the film) along the direction perpendicular to its surface. It is likely, that deformation is realized through the mechanism of radiation creep [13] rather than plastic deformation. The point is that at room temperature (where the most of experiments have been performed) ability of oxides to usual plastic deformation, as a rule, is rather low and the rate of atoms' displacement (in the process of their selective removing) in our experiments is very high and equals ~$10^{-2}$ dpa/s.

In addition, a special experiment confirms generation of stresses on the inter-phase film-substrate boundary in a course of irradiation when the transformation of insulator films of identical compositions into metal have been carried out for two different samples. The first one was the film without a substrate, and the second one was the film on a substrate. Study of electron diffraction of irradiated films (for the film on a substrate reflective diffraction was measured) shows that diffraction pattern for the film on a substrate was similar to that for stressed materials. Therefore, above-cited assumption concerning reasons leading to anisotropy of linear dimensions' variation with diminishing film volume by selective atoms removing seems to be quite probable.

All experiments with selective atoms removing out of insulators have been accompanied by essential variation of optical properties of films transformed into metals. Firstly, that transition leads to appearing characteristic metallic cluster on irradiated sections of films. To estimate variation of optical properties quantitatively, measurements of their absorption coefficient *S* have been carried out with the help of a micro-photometer.

In the course of those measurements, insulator film of a given thickness has been deposited on a glass substrate. Then absorption coefficients of that sandwich have been measured before irradiation and after selective atoms removing and transforming insulator into metal. For comparison, analogous measurements (but without irradiation) have been performed with films of various thickness produced by ion sputtering on a glass of the same metal.

Result obtained with oxide CoO which was produced by ion sputtering of bulk metal is a typical example which demonstrates variation of *S*-value. Initial absorption of insulator film CoO of 10 nm-thickness is characterised by the value $S=0$, that is the film is transparent. Comparison shows that optical transparency of the cobalt film (obtained by selective atoms removing out of cobalt oxide) with final thickness of ≈8 nm and specific resistivity $\rho \approx 1.3 \cdot 10^{-4}$ Ohm·cm corresponds to that of the pure cobalt film with close thickness of 7.5 nm. Hence, those measurements show that optical properties of metals produced by selective atoms removing out of insulators and pure metals produced by ion sputtering are in a good agreement.

## Perspectives and conclusions

In the present paper, physical principles are described and conditions are formulated allowing to remove selectively atoms of certain sort out of di- or polyatomic solids by means of their thin films or thin layers irradiation by accelerated particles. Noted are all essential features of selective atoms removing process following from the physical mechanism through which that process is realised. We demonstrated experimentally the possibility of such a selective atoms removing, confirmed the process mechanism and investigated its most essential features. In the course of experiments, we revealed that selective atoms removing out of di- and poly-component compounds is accompanied by radical variations of some most important physical properties of materials, such as electrical, magnetic and optical ones. Investigations show that the modification of materials' properties is a consequence of their atomic composition variations and those structural transformations (phase transitions) which accompany selective atoms removing out of materials.

Our experiments allow to outline basic phenomena accompanying selective atoms removing. However, many details of physical processes leading to observed radical variations of material physical properties remain beyond the scope of our work. First of all, it is associated with complexity and unusual character of those phenomena, which accompany selective atoms removing. In experiments performed, from 50 up to 75 % atoms have been removed out of material volume! That is why we have no detailed explanation of observed distinctions between properties of thin metal films produced by selective atoms removing and those produced by sputtering pure metals as well as distinctions between thin film properties and those of bulk materials.

However this circumstance, as it happens over and over again during science and technique history, is not a insuperable obstacle for practical utilisation of physical phenomena or processes which seem to be perspective for solving important practical problems. This is, undoubtedly, true in the case discussed. At the moment, ways are more or less clear that permit to reduce significantly distinctions between properties of thin metal films produced by the above-described transformation process and those of bulk materials. However, these items are far beyond the scope of the present paper which aim is, first of all, to demonstrate principally new opportunities given by the method of selective atoms removing out of di- or polyatomic compounds with the help of accelerated particle beams. It is evident that it is presently impossible to outline anywhere clear and completely the universe of potential compounds which are the best objects for selective atoms removing. In addition, it is clear that simultaneous removing atoms of several sorts is possible in polyatomic compounds. In doing so, relative rates of their removing could vary. We observed similar effects dealing with some polyatomic substances. All that evidences for the high potential and very wide application field of selective atoms removing technique to obtain local spatial variations of atomic constitution in thin films or layers.

To our mind, the opportunity to remove selectively atoms of a certain sort out of di- or polyatomic compounds is, as such, of great interest in respect to create future technologies associated primarily with needs of micro- and, more precisely, nanoelectronics as well as many other problems. Practical application perspectives of ion-beam atoms removing method are conditioned by a number of reasons:

1. Ion beams have numerous and important advantages:
   - effects of back-scattering inherent, e.g., for electron beams could be practically avoided that results in increasing spatial resolution of patterns on usual thin films deposited on massive substrates or thin layers on/in massive samples.
   - short wave lengths of incident particles important for high resolution could be obtained with low enough accelerating voltages.
2. The method could be used to create directly (avoiding lithography) needed spatial modulations of atomic constitution and physical properties of a material, say, metallic or semiconductive “patterns” in insulators, magnetic drawings in nonmagnetic substances, light guides in opaque media, etc.
3. Calculations show that the ion-projective system with proton beam parameters needed for the effective selective atoms removing could be created. Such a system (based on quintuple reducing of original mask pattern) could provide pattern spatial resolution of about 3 nm for 1 $cm^2$ area. It is important that upon termination of the process protons (hydrogen) could leave a material due to diffusion without any negative influence on material properties.
4. One-stage character and simplicity of selective atoms removing process open potential opportunity to create plain and cheap technology for obtaining multilayer bulk nanostructures of various applications. It is essential that in contrast to modern lithographic micro-technologies, here we could limit ourselves by two reiterating in each layer in-vacuum operations only - successive depositions of needed matrix films and drawing a given bulk picture (by selective atoms removing) with required spatial localisation of certain physical properties.

## Acknowledgements

Authors are thankful to their colleagues V. Ryl’kov, K. Prikhod’ko, A. Davidov, K.Maslakov, D. Kovalev and N. Chumakov for numerous useful discussions and help.

## Reference

1. Landau, L.D., Lifshitz, E.M., *Mechanics.* 3rd edition, Vol. **1**.,Butterworth-Heinemann,1976.
2. Tompson, M.W., *Defects and Radiation Damage in Metals.* Cambridge, at the University Press, 1969.
3. Klinger, et al., The Defect Creation in Solids by a Decay of Electronic Excitations. *Soviet Physics Uspekhi,* v.**28**, N 11, p.994 (1985)
4. *Handbook of Thin Film Technology,* eds. Leon I. Maisell and Reinhard Glang, McGraw Hill Hook company, 1970.
5. Hirsch, P.B., Howie, A., Nicholson, R.B., Pashley, D.W., Whelan, M.J., *Electron Microscopy of Thin Crystals*, London, Butterworths, 1965.
6. Wiesendanger, R., Guntherodt, H.-J., *Scanning Tunneling Microscopy II. (Further Application and Related Scanning Techniques.),* Springer-Verlag,Berlin,Heidenburg, New York, London, Paris, 1993.
7. *Practical Surface Analysis by Auger and X-rays Photoelectron Spectroscopy* eds. D.Briggs and M.P.Seah, John Wiley & Sons, Chichester, New-York, Brisbane, Toronto, Singapore, 1983.
8. Chopra, K.L., Randlett, M.R., Duff, R.N., Face-centered Cubic Modification in Sputtered Films of Tantalum, Molybdenum, Tungsten, Rhenium, Hafnium and Zirconium. *Phil. Mag.*, v. **16**, N 140, p. 261(August 1967).
9. Chopra, K.L., Randlett, M.R., Duff, R.N., Face-centered Cubic Tungsten Films Obtained by. *Appl. Phys. Lett.,* v.**9**, N11, p. 402, 1966.
10. Cahn, R.W., Metallic Glasses, *Contemporary physics,* v.**21**, N.1, p. 43, (1980).
11. Abrikosov, A.A., *Fundamentals of the Theory of Metals,* North-Holland, Amsterdam, 1988.
12. Altshuler, B.L., Aronov, A.G., in *Electron-Electron Interactions in Disoredered Systems,* eds. Efros, A.L. and Pollak, M., North-Holland, Amsterdam,1985.
13. *Physical Metallurgy,* 3rd edition, eds. Cahn R. W. and Haasen P., North-Holland Physics Publishing, 1983.

Typical variations of microstructure and diffraction patterns in the course of selective removing atoms of a given sort:

a) Dark-field image of a microstructure fragment for di-atomic compound after electron irradiation ($E$=200keV). It is seen non-simultaneity of structure transformations over the area of the irradiated part due to Gauss-like distribution of electron beam intensity. In the center, one could see crystallites of a new phase surrounded with an amorphous phase, while in the periphery, there are original phase crystallites not subjected to irradiation.

b) Combined diffraction pattern: top - from original polycrystal phase (before irradiation); bottom-from intermediate amorphous phase.

c) Combined diffraction pattern: top-from intermediate amorphous phase; bottom - from crystallites of a new phase formed in the course of irradiation;

d) Combined diffraction pattern: top-from crystallites of a new phase formed in the course of irradiation; bottom - from original polycrystal phase (before irradiation).

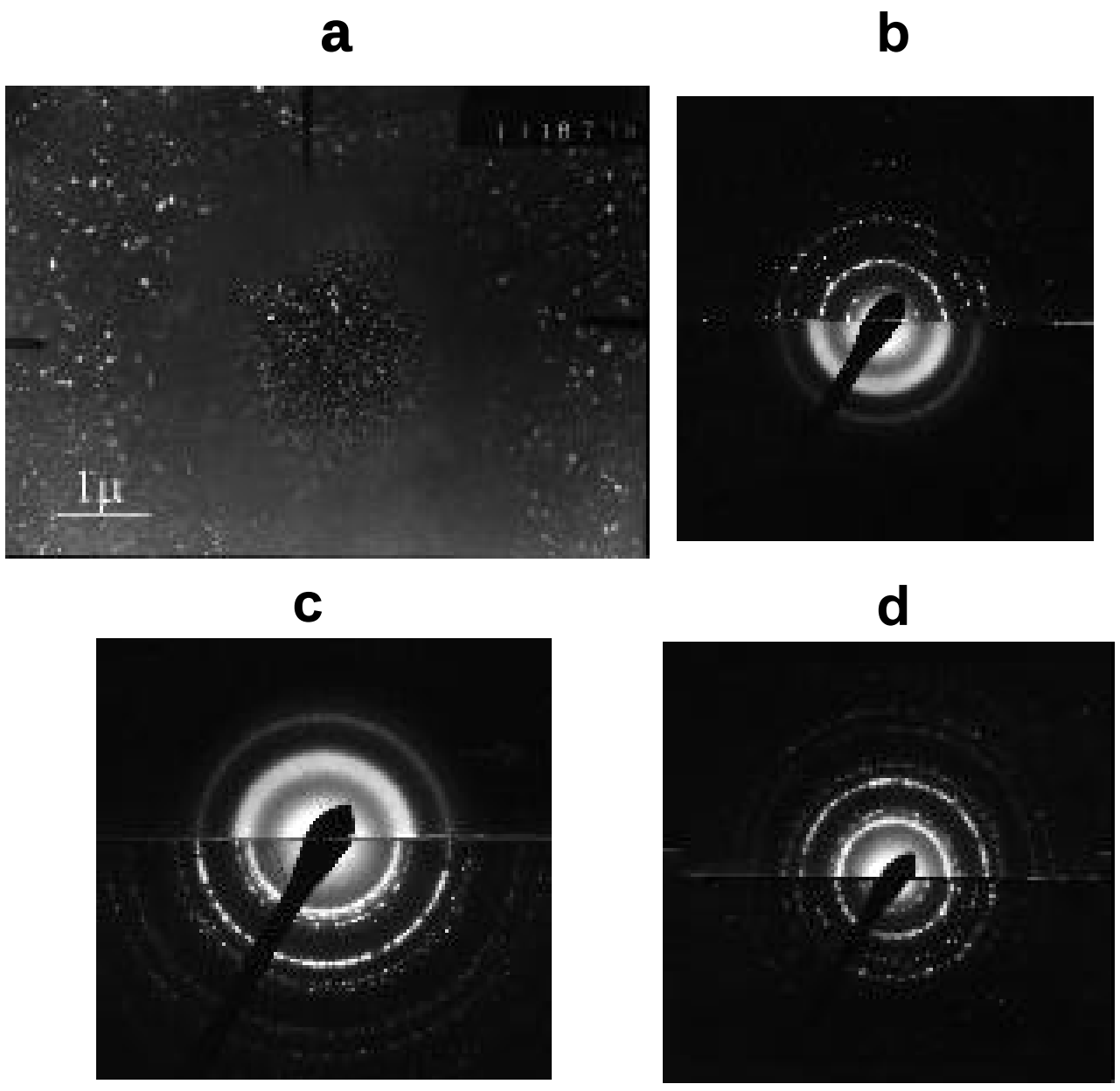

Fig. 1

Dose dependences of resistance for $WO_3$ film (100Å) under irradiation with different proton energy:
curve 1: $E$=1000eV;
curve 2: $E$=1050eV, $E_{max}$-$E_{d2}$≅2eV;
curve 3: $E$=5000eV, $E_{max}$-$E_{d2}$≅85eV.

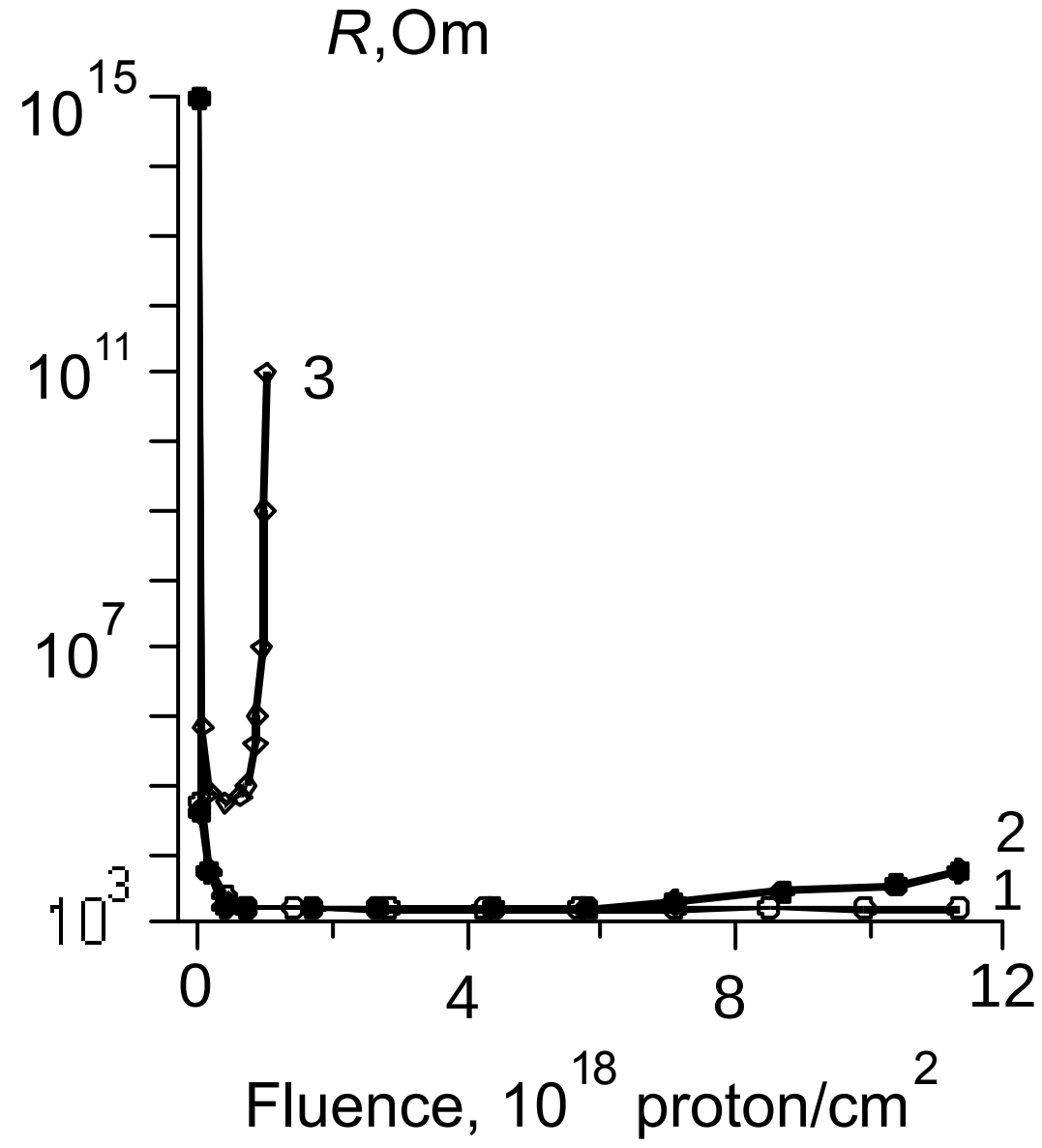

Fig.2

Dose dependence of specific resistivity for CuO - film (100 Å) under proton irradiation ($E$=450 eV).

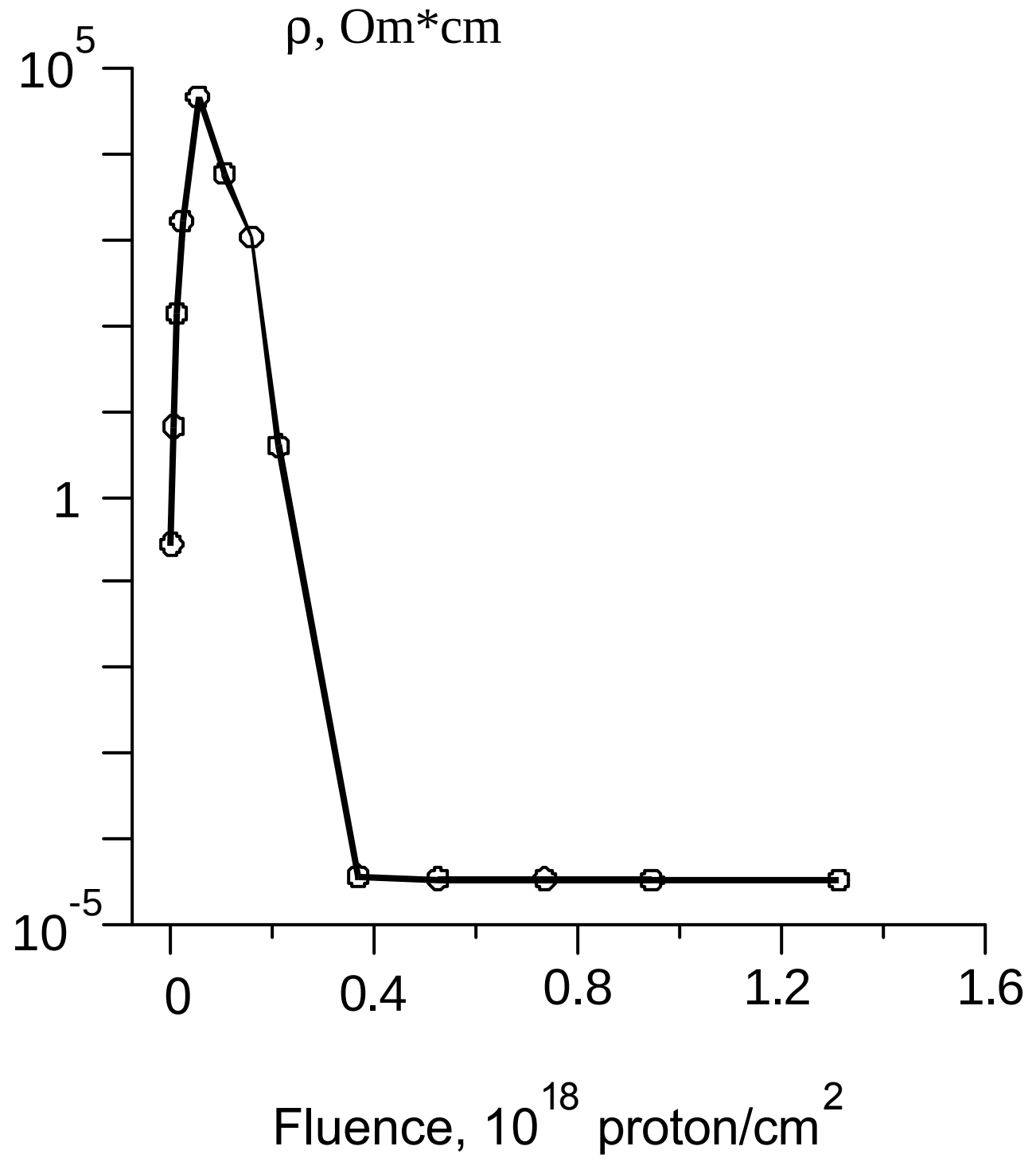

Fig.3

X-ray photoelectron spectra for $WO_3$-films (100 Å) in the initial state and after proton irradiation of various dose ($E$=1000 eV).
curve 1 : before irradiation;
curve 2 : dose=$7 \cdot 10^{16}$ proton/$cm^2$;
curve 3 : dose=$7 \cdot 10^{17}$ proton/$cm^2$.

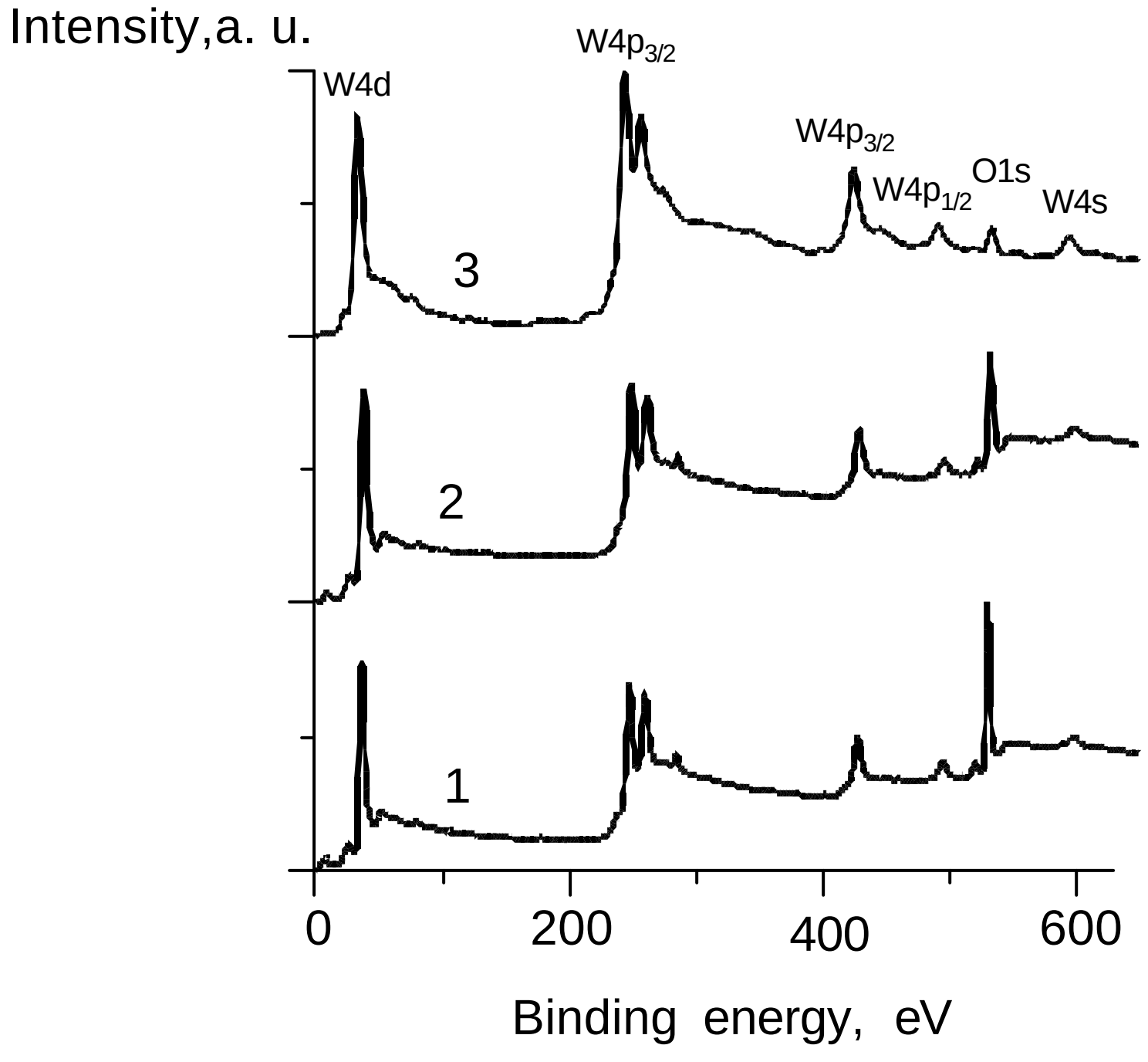

Fig.4

Temperature dependences of reduced resistivity *R*(T)/*R*(300 K) for films (100Å):

curve 1: CoO-oxide before irradiation;

curve 2: CoO-oxide after irradiation up to dose $0.5 \cdot 10^{18}$ proton/cm$^2$;

curve 3: CoO-oxide after irradiation up to dose $10 \cdot 10^{18}$ proton/cm$^2$;

curve 4: Co obtained by ion sputtering of pure metal.

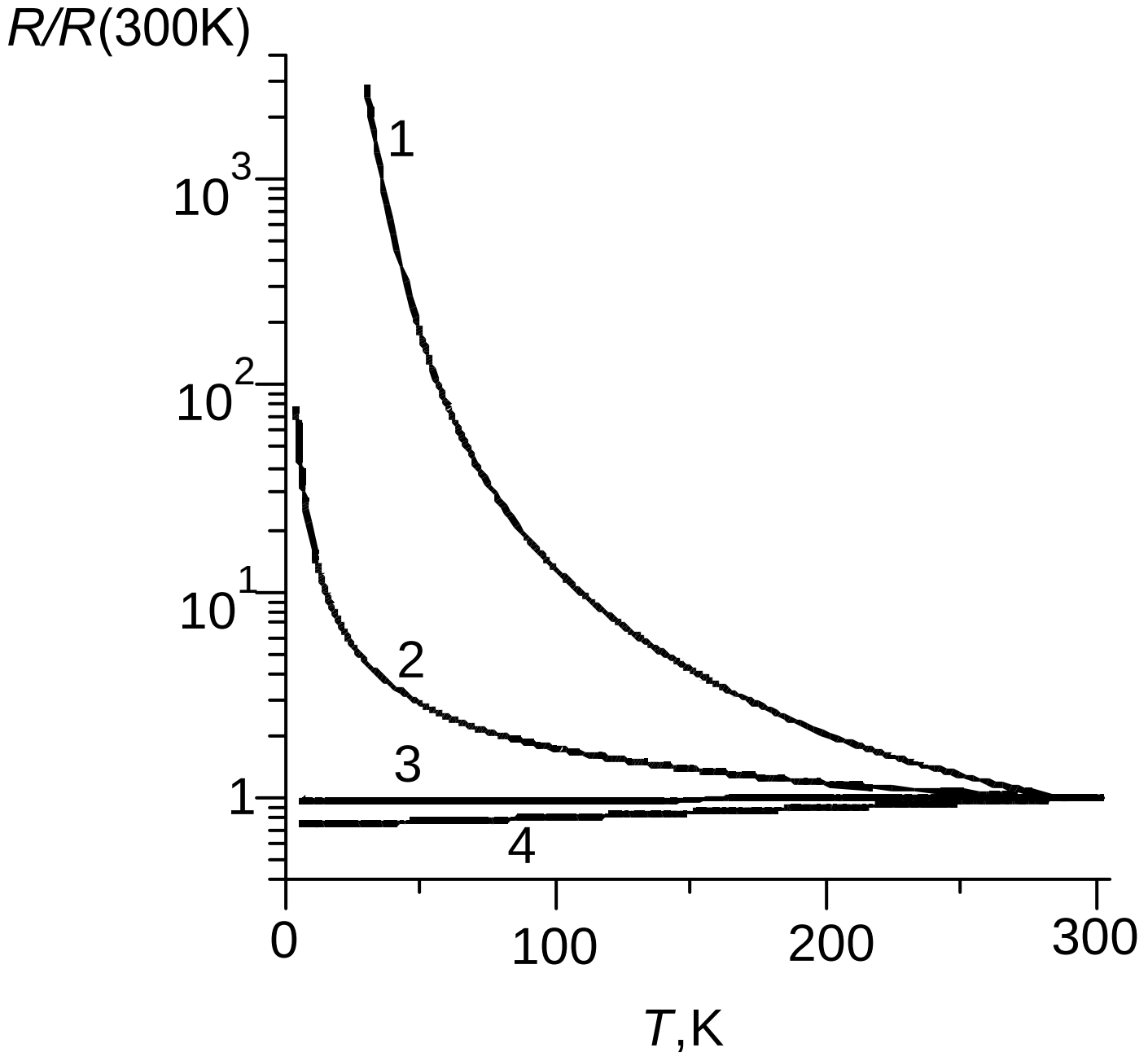

Fig.5

Temperature dependences of reduced resistivity *R*(T)/*R*(300 K) for films (100 Å):
curve 1: Co produced by selective atoms removimg out of oxide;
curve 2: Cu produced by selective atoms removimg out of oxide;
curve 3: Co obtained by ion sputtering of pure metal;
curve 4: Cu obtained by ion sputtering of pure metal.
Curves for pure Co (3) and Co produced from oxide (1) is stressed in insets which demonstrate the existence of minimums at low temperatures for Co curves at enlarged scale.

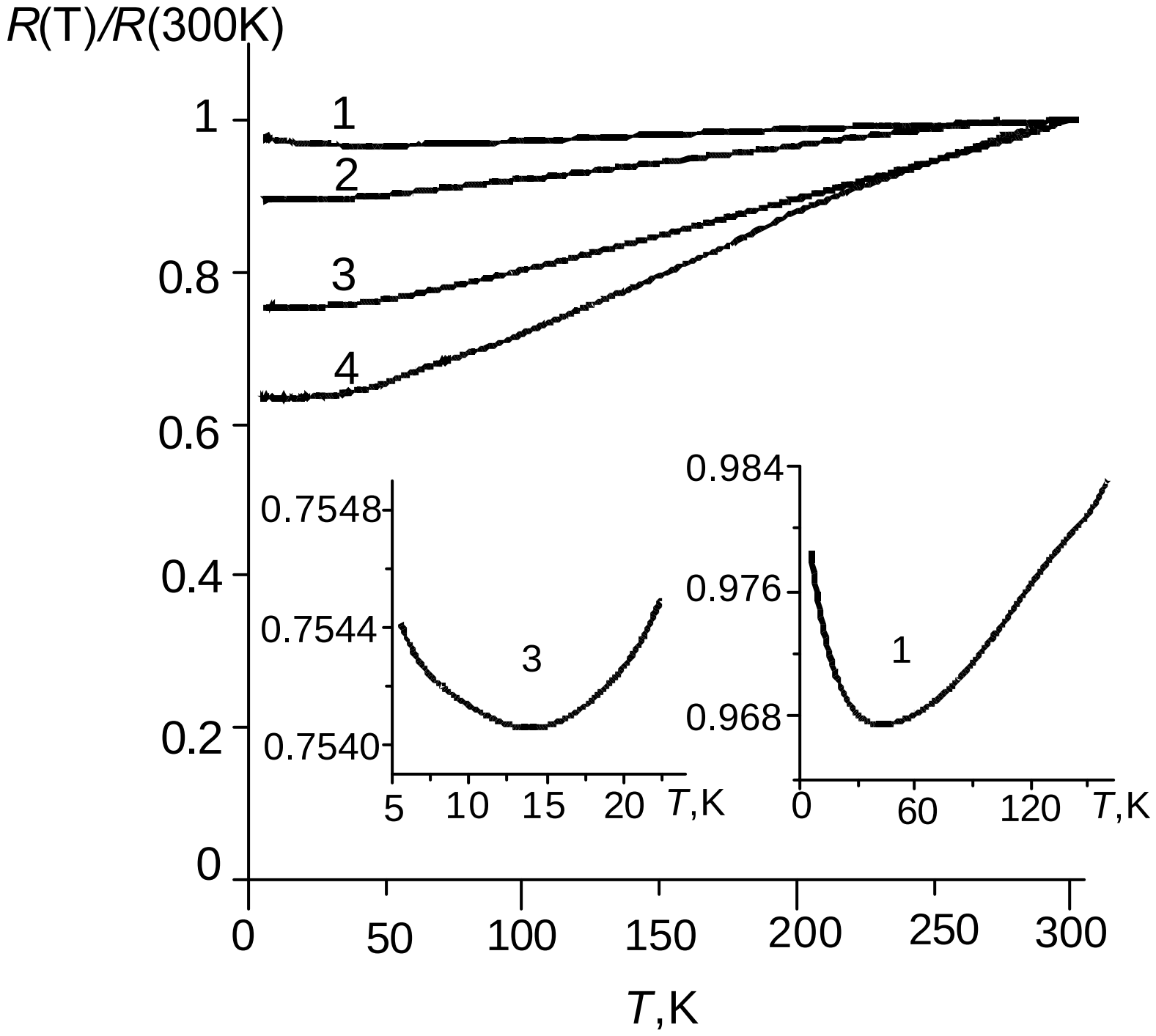

Fig.6

Specific resistivity dose dependences of the sandwich:
film $WO_3$ (100 Å) - carbon under proton irradiation ($E$=900 eV):
curve 1: additional carbon layer thickness is 50 Å;
curve 2: additional carbon layer thickness is 400 Å.

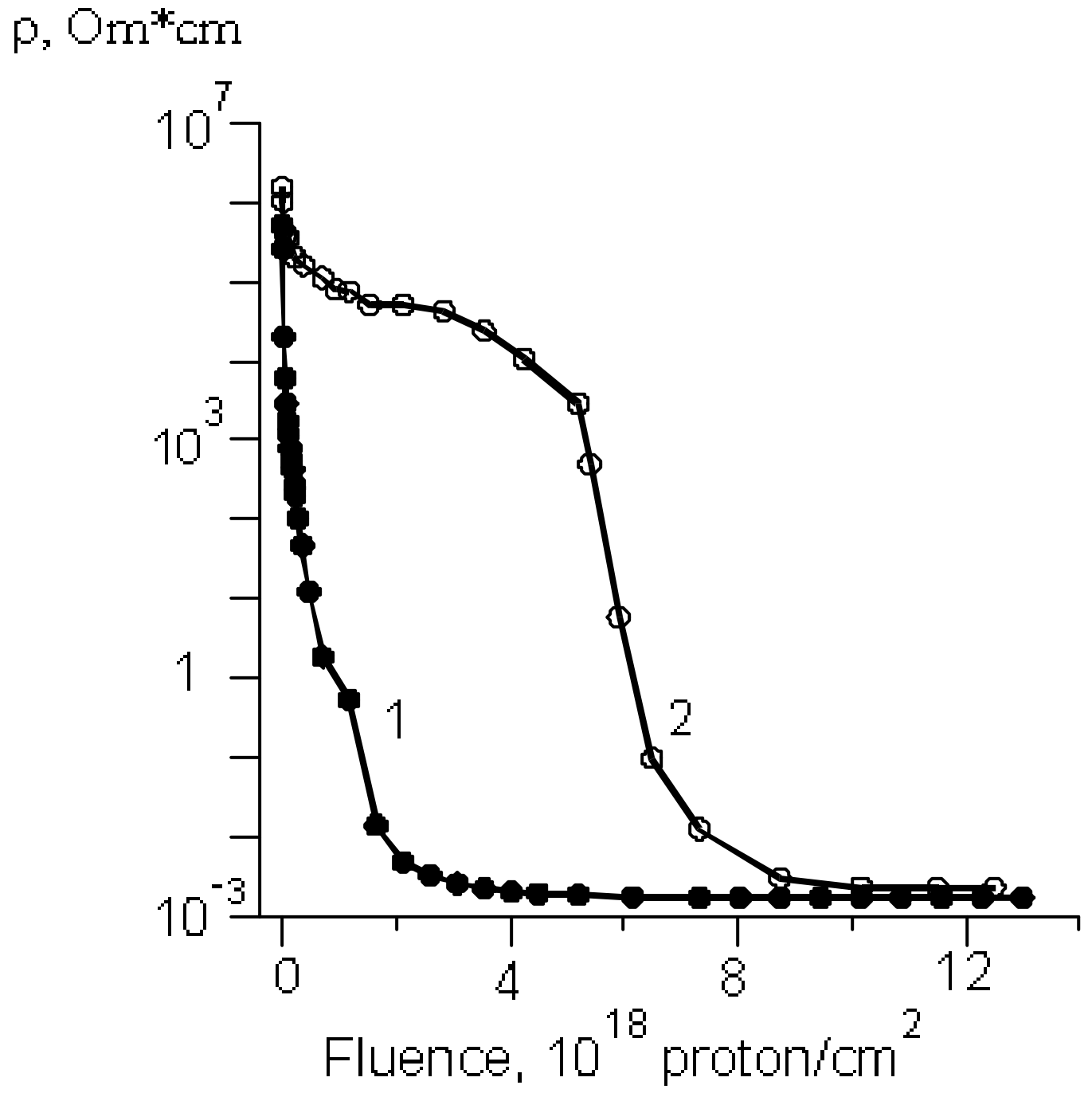

Fig.7

Magnetization curves for original insulator films and metals produced out of them by selective atoms removing:
curve 1: CoO - film (500 Å) before irradiation;
curve 2: Co-film produced by proton irradiation ($E$=600eV) of CoO film (500 Å);
curve 3: Fe-Co-V-O film (100 Å) before irradiation;
curve 4: Fe-Co-V-film produced by proton irradiation ($E$=400eV) of Fe-Co-V-O film (100 Å).

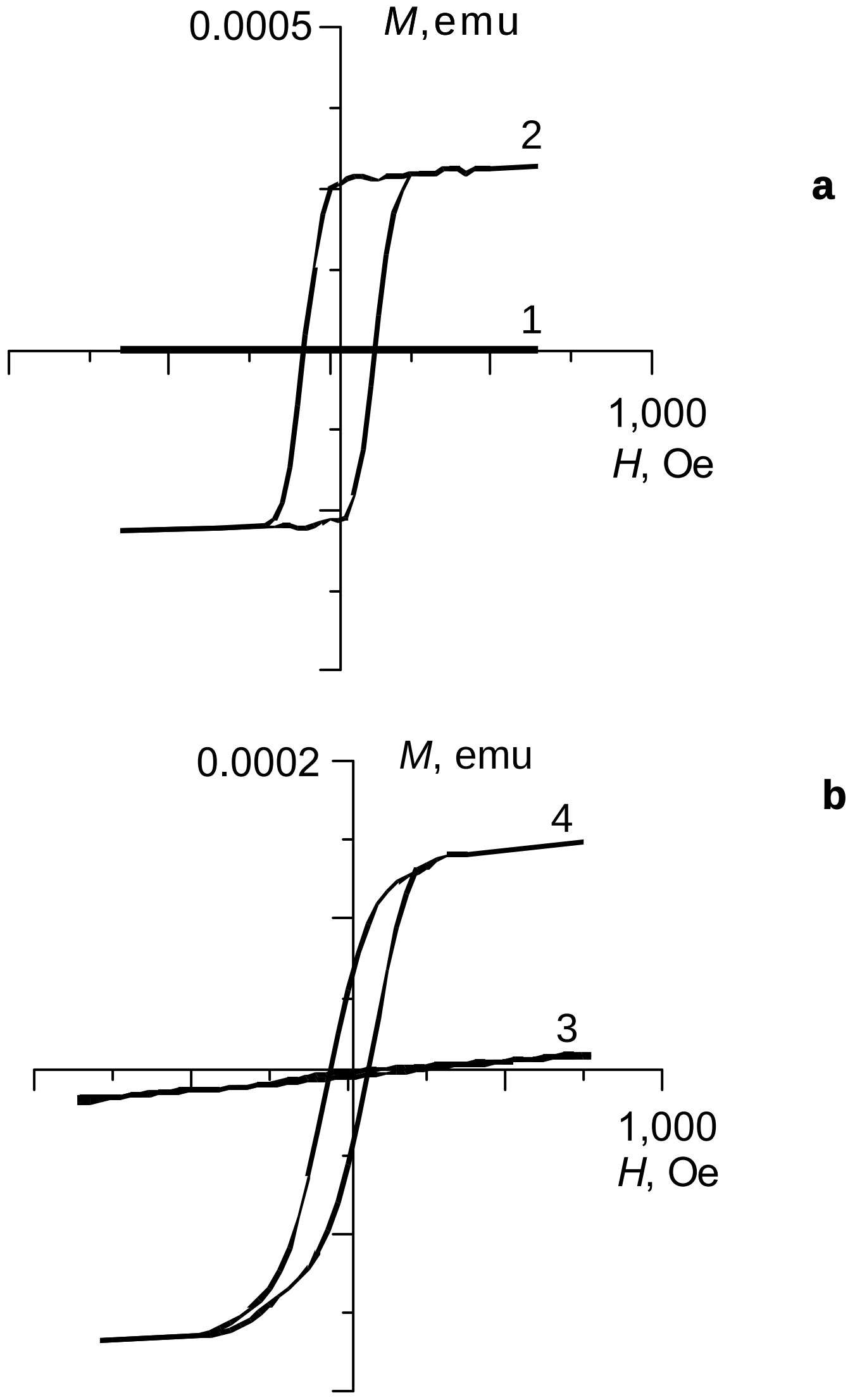

Fig. 8

Dose dependences of specific resistivity for ferromagnetic metal oxide films:

curve 1: CoO - film(500 Å) under proton irradiation ($E$=600eV);

curve 2: Fe-Co-V-O - film (100 Å) under proton irradiation ($E$=400eV).

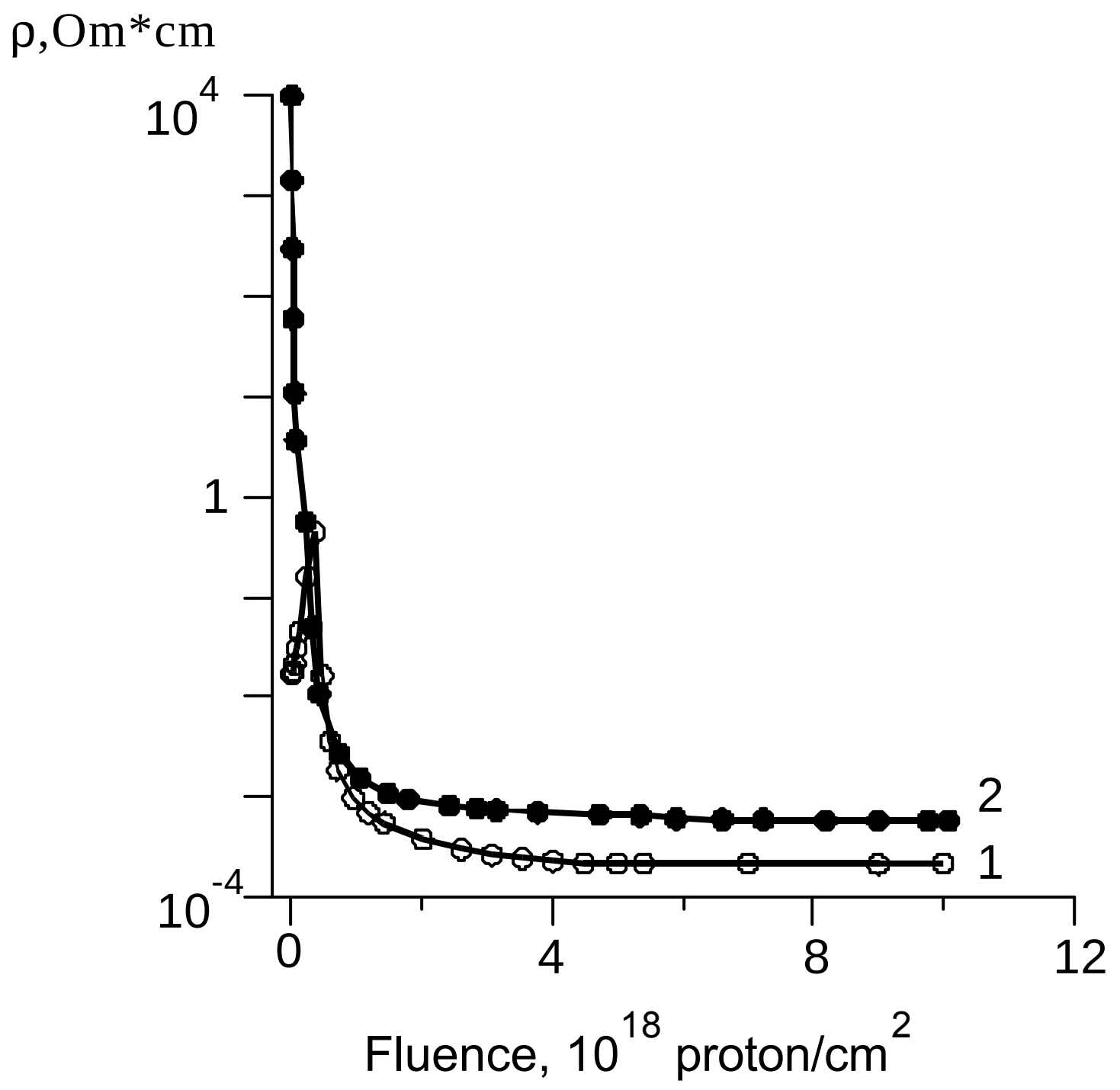

Fig.9